\documentclass[aps,onecolumn,superscriptaddress,floatfix,pt14,color]{revtex4}

\usepackage{graphicx}
\usepackage{epsfig}
\usepackage{amssymb}

\usepackage[dvips]{color}
\definecolor{Red}{rgb}{1.00, 0.00, 0.00}

\begin{document}
\newcommand{\be}{\begin{eqnarray}}
\newcommand{\ee}{\end{eqnarray}}
\newcommand\del{\partial}
\newcommand\nn{\nonumber}
\newcommand{\Tr}{{\rm Tr}}
\newcommand{\mat}{\left ( \begin{array}{cc}}
\newcommand{\emat}{\end{array} \right )}
\newcommand{\vect}{\left ( \begin{array}{c}}
\newcommand{\evect}{\end{array} \right )}
\newcommand{\tr}{\rm Tr}
\def\conj#1{{{#1}^{*}}}
\newcommand\hatmu{\hat{\mu}}
\newcommand\noi{\noindent}
\newcommand{\Ree}{{\rm Re}}
\newcommand{\Imm}{{\rm Im}}

\voffset 0.5cm

\title{The Fluctuations of the Quark Number and of the Chiral Condensate}

\author{M.P. Lombardo}
\affiliation{INFN-Laboratori Nazionali di Frascati, I-00044, Frascati (RM), Italy} 
\author{K. Splittorff}
\affiliation{The Niels Bohr Institute, Blegdamsvej 17, DK-2100, Copenhagen
  {\O}, Denmark} 
\author{J.J.M. Verbaarschot}
\affiliation{Department of Physics and Astronomy, SUNY, Stony Brook,
 New York 11794, USA}

\date   {\today}
\begin  {abstract}

The distributions of the quark number and chiral condensate over the
gauge fields are computed for QCD in Euclidean space at nonzero quark 
chemical potential. As both
operators are non-hermitian the distributions are in the complex
plane. Moreover, because of the sign problem, the distributions are
not real and positive. The computations are carried
out within leading order chiral perturbation theory and give a
direct insight into the delicate cancellations that
take place in contributions to the total baryon number and the chiral
condensate.

\end{abstract}

\maketitle
\newpage
 
\section{Introduction}

The phase diagram of strongly interacting matter is determined by the
behavior of the chiral condensate, $\langle\sigma\rangle$, and
the quark number, $\langle n \rangle$. These quantities signal the 
breakdown of chiral symmetry and the formation of baryonic
matter. Not only the expectation values of $\sigma$ and
$n$ are of interest, but also their fluctuations
are central to our understanding of strongly interacting matter
since they may be visible in
fluctuations of the observables measured in heavy ion collisions
\cite{SRS1,SRS2,JK,misha-fluc}.    

In Euclidean space both $\sigma$ and $n$ are complex when evaluated 
at nonzero chemical potential
for a typical gauge field background. 
The reason why $n$ and $\sigma$ take complex values is that the
fermion determinant at nonzero chemical potential is complex: if we
write $\det(D+\mu\gamma_0+m)=r\exp(i\theta)$ we have
\be
\label{1}
n & \equiv &  
 \frac{d}{d\mu} \log\det(D+\mu\gamma_0+m) 
=  \frac{d}{d\mu} \log
 r + i\frac{d}{d\mu} \theta  \\
\sigma & \equiv &  
\frac{d}{d m } \log\det(D+\mu\gamma_0+m) 
=  \frac{d}{d m} \log
 r + i\frac{d}{d m} \theta.  \nn
\ee
The fluctuations, therefore, take place in
the complex  $\sigma$- and $n$-plane. Since the fermion determinant is
also complex valued for non zero chemical potential, fluctuations in all
directions of the complex $\sigma$- and $n$-plane contribute
to the expectation values $\langle\sigma\rangle$ and
$\langle n\rangle$, which of
course are real. A description of fluctuations of
$\sigma$ and $n$ is contained in their distributions, $P_{\sigma}$ and
$P_{n}$. Here we derive these distributions analytically to leading
nontrivial order in chiral perturbation theory. 
From (\ref{1}) we see that the imaginary part of the quark number
operator is directly 
related to the phase of the fermion determinant. We therefore expect 
that the cancellations due to fluctuations of the phase are tightly
linked to the distribution of the quark number in the imaginary direction. 
The results presented below confirm this expectation.

Chiral perturbation theory \cite{CPT} is the low energy
limit of QCD which describes the strongly interacting theory in terms of 
weakly interacting Goldstone modes corresponding to the spontaneous
breakdown of chiral symmetry. Since the pions have zero quark
charge, the expectation value of the quark number in chiral
perturbation theory is automatically zero. The square of the quark
number
\be
\label{2}
\langle n_q^2 \rangle = \frac{1}{Z} \ \frac{d^2}{d\mu^2}  \ Z 
= \langle n^2 \rangle + \langle\left(\frac{d n}{d\mu}\right) \rangle 
\ee
is of course also zero since in chiral perturbation theory the
partition function, $Z$, is independent of $\mu$.
The distribution of the values which the 
quark number operator takes as the gauge fields fluctuate is, 
however, nontrivial even when evaluated within chiral perturbation
theory. To see that this is necessarily the
case, let us consider the second moment of $n$ 
i.e.~$\langle n^2\rangle = \int dx dy (x+iy)^2 P_{n}(x+iy)$. 
If the second moment is non zero the 
distribution $P_n$ is nontrivial, i.e. not a two dimensional
$\delta$-function at the origin of the complex $n$-plane.  
 
Let us emphasize that $\langle n^2\rangle$ is not the average of the
square of the quark number, see (\ref{2}), rather the second moment
must be thought of as    
\be
\label{3}
\langle n^2 \rangle=\frac{1}{Z} \ \frac{d}{d\mu_u}\frac{d}{d\mu_d} \  Z
\ee
evaluated at degenerate chemical potentials and quark masses for the
up and down quark. Even though there is an equal number of quarks and
anti-quarks associated with each pion, the vev of 
 $\langle n^2\rangle$, is nonzero if there are
correlations between 
the quarks and the anti-quarks of different flavor. Confinement of
quarks and anti-quarks into pions strongly suggests that such a
correlation exists in chiral perturbation theory, and in fact $\langle
n^2 \rangle$ takes a non zero value even at $\mu=0$, see
for example \cite{LSV}. 
At zero chemical potential $\langle n^2 \rangle$ coincides with the
off diagonal quark number susceptibility   
which has been computed analytically by high temperature
perturbation theory 
\cite{Blaizot:2001vr} as well as by numerical lattice simulations
\cite{susclat}. 

Since this is a likely source of confusion let us again stress that $P_n$
gives the distribution of $n$, defined in (\ref{1}), over the ensemble
of gauge fields: If one makes a frequency plot of the values obtained
for $n$ in a lattice QCD ensemble of gauge fields the shape which
emerges is described by $P_n$. The first moment measured on this
distribution gives the average quark number. The second moment
measured on this distribution, however, only give the first of the 
two terms which combine to give the square of the quark number, cf. 
(\ref{2}). If we phrase this in terms of the eigenvalues $z_k$ of $\gamma_0
(D+m)$ we have 
\be
n_q & = &  n = \sum_k \frac 1{z_k+\mu} \\
n_q^2 & = & \sum_{k \ne l} \frac 1{z_k+\mu} \frac
1{z_l+\mu} \nn\\
n^2 &  = &  \sum_{k , l} \frac 1{z_k+\mu} \frac
1{z_l+\mu} = [\sum_k \frac 1{z_k+\mu} ]^2 . \nn
\ee
So $\langle n_q^2 \rangle$ is not the average of a square and
consequently not the second moment of a distribution over the gauge
fields. The reason is that $n_q^2$ contains
correlations due to the Pauli principle.

On a more technical level, the reason
for the $\mu$ dependence of $P_n(x+iy)$ can bee seen from the
generating function 
of the distributions. These involve quarks with different 
chemical potentials, i.e. the isospin chemical potential,
$\mu_1-\mu_2$, is nonzero and
couples non-trivially in the generating
functions. This point will be explained explicitly below. 
The success of chiral perturbation theory in predicting the
behavior of QCD at nonzero chemical potential has been demonstrated   
clearly by recent comparisons to lattice QCD results both in the
$p$-regime \cite{Simon,SV-thermo,D'Elia,Gattringer} and the
$\epsilon$-regime \cite{DHSS,DHSST,BW,SplitSvet,ABSW}.

The phase of the fermion determinant may lead to exponentially large
cancellations in the computation of expectation values. If we would try to
measure this expectation value numerically with lattice QCD simulations, 
we would be confronted with
 numerical errors which are exponentially hard to handle. This
is {\sl the QCD sign problem} and severely limits first principle
studies of the QCD phase diagram (see \cite{reviews} for recent
reviews of the QCD sign problem).   

The results for the distributions that are derived below give a 
direct insight in the cancellations caused by the sign problem. For
example, the vanishing value of the baryon number (within chiral
perturbation theory) is obtained only
after a delicate cancellation between the contribution from $\Ree [n]$ and
from $\Imm[n]$. The range of the fluctuations in the
complex quark number plane which must be taken into account, in order
to obtain the total baryon number, grows like the four volume. In
contrast, the width of the distribution for 
the ensemble where the
phase of the fermion determinant is ignored (the phase quenched
ensemble) only scales like the square root of the volume.   
The numerical challenge faced by lattice QCD is to go from the phase
quenched (simulation) ensemble to the full (target) ensemble. One
therefore needs to sample the far tails of the distributions.   
The analytic insight we have obtained here can in this way help to 
understand the limitations of present lattice gauge simulations at non
zero chemical potential and give hints for future developments. 
Besides the reweighting method \cite{Glasgow,fodor1,fodor2}
discussed here, the results obtained are also relevant for the 
Complex Langevin method \cite{Parisi,KW,FOC,AFP,aarts} as well as for the
density of states method \cite{AN,AANV,Schmidt,Ejiri}.

Throughout the paper we will work with two flavors. We refer to the
this theory as the $N_f=1+1$ theory. The phase quenched theory is
referred to as $1+1^*$ since the absolute square of the fermion
determinant corresponds to a quark and a conjugate quark.

This paper is organized as follows. First we consider the region where
$2\mu$ is less than the pion mass and the mean fields do not depend 
on the chemical potential.
We 
derive the distribution of the quark number operator as well as that
of the chiral condensate. Results are given both for
QCD with dynamical quarks and for the partially quenched case. Implications 
for lattice QCD are discussed in section
\ref{sec:numproblem}.  
Before concluding we discuss the distribution of the
quark number operator and chiral condensate for larger values of the
chemical potential. The distribution of the quark number for imaginary
chemical potential is evaluated in the Appendix.

\section{General definitions and known results from CPT}

Before starting  the actual computation of the distributions, in this section
we introduce notations and explain why chiral
perturbation theory can give information about QCD at nonzero quark chemical
potential. 
 
\hspace{3mm}

The quark number operator is the logarithmic derivative of the
fermion determinant with respect to the quark chemical potential
$\mu = \mu_1 + \mu_2$ 
\be\label{ndef}
n(\mu)\equiv \Tr\,\frac{\gamma_0}{D+\mu\gamma_0+m},
\ee
while for $\sigma$ we differentiate with respect to the
quark mass (we consider degenerate flavors)
\be
\sigma(\mu) \equiv \Tr\frac{1}{D+\mu\gamma_0+m}.
\ee 
At low temperatures, the QCD partition function and its low energy limit in
the form of a chiral Lagrangian are
independent of $\mu$
until the chemical potential is sufficient to balance the energy required to
create a baryon. 
Since the expectation value of the quark number and the 
chiral condensate 
are derivatives of the partition
function with respect to $\mu$, 
they are independent of $\mu$ when evaluated in chiral
perturbation theory.
 On the contrary,
expectation values that cannot be written as derivatives of the 
partition function  with respect to $\mu$ 
may still depend on the chemical
potential.
This happens when the generating function for the
operator under consideration includes quarks with different values of the
chemical potential. 

To illustrate this let us compute the expectation
value of $n^2$, the cross correlation introduced above.  
In order to obtain $n^2$ we
start from the generating function 
\be
Z_{1+1}(\mu,\mu_1) = 
\langle {\det}(D+\mu \gamma_0 +m){\det}(D+\mu_1 \gamma_0+m) \rangle  
\ee
and differentiate with respect to the two different chemical potentials
\be
\langle n^2 \rangle_{1+1} 
= \frac{1}{Z_{1+1}(\mu,\mu)} \lim_{\mu_1\to\mu} 
\frac{d}{d\mu}\frac{d}{d\mu_1} Z(\mu,\mu_1).
\ee
Since the chemical potentials in the generating function are different,
there is a nonzero component of the isospin chemical
potential, $\mu-\mu_1$. It is this component that induces a nontrivial
chemical potential dependence in chiral perturbation theory and hence a nonzero
result after differentiation with respect to
$\mu$ and $\mu_1$.   
Since the free energy is an even function of the isospin
chemical potential (for degenerate quark masses)
it takes two derivatives to obtain a nonzero value
for $\mu_1\to\mu$. 

To one-loop order in chiral perturbation theory we have  
\be
\frac{Z_{1+1}(\mu,\mu_1)}{Z_{1+1}(\mu,\mu)} 
= e^{G_0(\mu,\mu_1)-G_0(\mu=0)},
\ee
where the one-loop free energy is (the divergent part of $G_0$ is
independent of $\mu$ and does not contribute)
\be 
G_0(\mu,-\tilde\mu)
& = & \frac{Vm_\pi^2T^2}{\pi^2}\sum_{n=1}^\infty 
\frac{K_2(\frac{m_\pi n}{T})}{n^2}\cosh(\frac{(\mu+\tilde\mu) n}{T}).   
\label{G0}
\ee 
Thus 
\be
\langle n^2 \rangle_{1+1} 
= \lim_{\mu_1\to\mu} \frac{d}{d\mu}\frac{d}{d\mu_1} G_0(\mu,\mu_1).
\label{nbSQ}
\ee
This quantity is usually referred to as the off diagonal quark number
susceptibility and is denoted by $\chi_{ud}^B$. 

Below we will use the
notation $\Delta G_0(\mu_1,\mu_2)=G_0(\mu_1,\mu_2)-G_0(0,0)$ and 
\be 
\label{notation-n}
\nu_I &\equiv& \left . 
\frac d {d\mu_1} \Delta G_0(\mu_1, -\mu) 
\right|_{\mu_1 = \mu},\\
\chi^B_{ud} &\equiv& \left .
 \frac {d^2} {d\mu_1d\mu_2} \Delta G_0(\mu_1, \mu_2)
 \right |_{\mu_1 = \mu_2 =\mu}\nn,\\
\chi^I_{ud} &\equiv& \left . 
 \frac {d^2} {d\mu_1d\mu_2} \Delta G_0(-\mu_1, \mu_2)
 \right |_{\mu_1 = \mu_2 =\mu},\nn
\ee
for the $\mu$ derivatives of the free energy and 
\be
\label{notation-m}
\delta\Sigma^B & = & \left. \frac{d}{d\tilde{m}}\Delta
G_0(\mu,\mu,m,\tilde{m})\right|_{\tilde{m}=m}, \\
\delta\Sigma^I & = & \left. \frac{d}{d\tilde{m}}\Delta
G_0(\mu,-\mu,m,\tilde{m})\right|_{\tilde{m}=m}, \nn \\
\chi_S^B & = & \left. \frac{d^2}{d\tilde{m}dm}\Delta
G_0(\mu,\mu,m,\tilde{m})\right|_{\tilde{m}=m}, \nn \\
\chi_S^I & = & \left. \frac{d^2}{d\tilde{m}dm}\Delta
G_0(\mu,-\mu,m,\tilde{m})\right|_{\tilde{m}=m}, \nn
\ee
for the one-loop contributions to the chiral condensate and
chiral susceptibility. Also for the chiral susceptibility
it is understood that we will only consider the off diagonal
component. 
Note that all of these quantities are extensive. For quantities labeled
by  $B$ we have that $\mu_q = \mu$ and $\mu_I =0$, and for the quantities
labeled with $I$ $\mu_q = 0$  and $\mu_I = \mu$. The quark mass dependence of 
$G_0$ is through $m_\pi$ via the Gell-Mann--Oakes--Renner relation,
\be
m_\pi^2 = \frac{\Sigma(m+\tilde m)}{F^2}.
\ee
Let us also stress
that all quantities with superscript $B$ are independent of the
chemical potential (in 1-loop chiral perturbation theory).

In table \ref{tab:1} we give  results for one-loop chiral
perturbation theory valid for $\mu<m_\pi/2$. The label $PQ$ refers to
the result obtained in a partially quenched ensemble. By definition
this ensemble is generated with the absolute value of the fermion
determinant but we keep the $\mu$ dependence of $n$ and
$\sigma$ as in the ordinary theory. We will use each of
these results to check the distributions of the quark number
operator and the chiral condensate.

\begin{table}[h!]
\begin{tabular}{|c|c|c|}
\hline
& ${\cal E} = 1+1$ & ${\cal E} = PQ$ \\[0.2cm] 
\hline
$\langle n \rangle_{\cal E}$ & 0 & $\nu_I$ \\[0.2cm]
$\langle n^2\rangle_{\cal E}$ & $\chi_{ud}^B$ & $\nu_I^2  + \chi^B_{ud}$\\[0.2cm]
$|\langle \sigma \rangle_{\cal E}|$ &  $\Sigma +2 \delta\Sigma^B $
&  $\Sigma + \delta\Sigma^I+\delta\Sigma^B $ \\[0.2cm]
$\langle \sigma^2 \rangle_{\cal E} $ 
& $\chi_S^B+(\Sigma +2 \delta\Sigma^B)^2 $ 
& $\chi_S^B+(\Sigma + \delta\Sigma^B +\delta \Sigma^I)^2$\\[0.2cm]
\hline
\end{tabular}
\caption{\label{tab:1} The first and second moment of $n$ and
  $\sigma$ to 
  one-loop order in chiral perturbation theory for $\mu<m_\pi/2$. The
  absolute value of the mean field result for the chiral condensate is
  denoted by $\Sigma$. In the partially quenched 
  (PQ) ensemble the operators of the full theory are evaluated for a
  gauge field background where the phase of the fermion determinant is
  ignored.} 
\end{table}

\section{The distribution of the quark number operator}

In this section we compute the distribution of the quark number in 
the 1+1 theory.
Since 
\be\label{nBherm}
n(\mu)^*=\left(\Tr\frac{\gamma_0}{D+\mu\gamma_0+m}\right)^*
=-\Tr\frac{\gamma_0}{D-\mu\gamma_0+m}=-n(-\mu)
\ee
the quark number operator is in general complex (it is purely
imaginary at $\mu=0$). 
The fluctuations of the quark number thus occur in the complex plane. We
first derive the distribution of the real part, $(n(\mu)-n(-\mu))/2$,
and of the  
imaginary part, $(n(\mu)+n(-\mu))/2$, of the
quark number. Then, finally, we compute the full distribution in the complex
quark number plane. As we shall see the distribution in the complex plane
factorizes into the distribution of the real part and the distribution
of the imaginary part.

\subsection{The distribution of the real part of the quark number}  
\label{subsec:PRenB1p1}

Here we derive the distribution of the real part of the quark number 
defined by  
\be
P^{1+1}_{\Ree[n]}(x)\equiv\left\langle
\delta\left(x-\frac{1}{2}(n(\mu)-n(-\mu))\right)\right\rangle_{1+1},
\ee
within one-loop chiral perturbation theory for $\mu<m_\pi/2$. 

\vspace{5mm}

First we represent the $\delta$-function as an integral
\be
\label{kint}
P^{1+1}_{\Ree[n]}(x) = \frac{1}{2\pi}\int_{-\infty}^\infty dk \
e^{-ixk}\left\langle e^{+i\frac{k}{2}(n(\mu)-n(-\mu))}\right\rangle_{1+1}.
\ee
Then we expand the exponential of the trace 
\be
\label{expandTr}
\left\langle e^{i\frac{k}{2} (n(\mu)-n(-\mu)) }\right\rangle_{1+1}
&=& \sum_{j=0}^\infty \frac{(ik/2)^j}{j!}
\left\langle\left(n(\mu)-n(-\mu)\right)^j\right\rangle_{1+1}.  
\ee
This shows that a probability distribution is determined by its
moments which we will compute next.

In order to compute the trace to the $j$th power we need to introduce $2j$
replica quarks (see \cite{replica} for an introduction to the replica
trick in chiral perturbation theory) 
\be
&&\hspace{-2cm}\left\langle\left(n(\mu)-n(-\mu)\right)^j\right\rangle_{1+1} \\
&=& \left  .
\lim_{n_i\to0}\frac{1}{n_1\cdots n_j}d_{\mu_1}\cdots d_{\mu_j} 
\left\langle\prod_{i=1}^j
\det(D+\mu_i\gamma_0+m)^{n_i}\det(D-\mu_i\gamma_0+m)^{n_i}\right\rangle_{1+1} 
\right |_{\mu_i=\mu}.\nn 
\ee 

In one-loop chiral perturbation theory the replicated generating function
for the real part  is given by  (this is where the assumption $\mu<m_\pi/2$
enters: for $\mu>m_\pi/2$ the generating
function is in a Bose condensed phase, see section \ref{sec:mpi>2mu})
\be
&&\hspace{-2cm}\left\langle\prod_{i=1}^j
  \det(D+\mu_i\gamma_0+m)^{n_i}(D-\mu_i\gamma_0+m)^{n_i} \right\rangle_{1+1} \nn\\
&=& \exp\left(\sum_{l\leq m=1}^j 2 n_ln_m (G_0(\mu_l,\mu_m)+G_0(-\mu_l,\mu_m))+\sum_{l=1}^j 2n_l(G_0(\mu_l,\mu)+G_0(-\mu_l,\mu)) \right).
\label{genFrenBdist}
\ee
We now take $d_{\mu_1}\cdots d_{\mu_j}$ of the generating function,
evaluate it at $\mu_i=\mu$ and take the replica limits $n_i\to0$. Note that
the term linear in the $n_k$'s includes $G_0(-\mu_l,\mu)$, and the derivative
with respect to~$\mu_l$ does {\sl not} vanish at $\mu_l=\mu$. Therefore 
terms with
even as well as odd values of $j$ contribute when we
evaluate the derivative $d_{\mu_1}\cdots d_{\mu_j}$ at $\mu_l=\mu$. All terms
with the same number of pairs from the first sum in (\ref{genFrenBdist})
give the same contribution to the $j$th moment. The combinatorial factor
for choosing $b$ pairs out of $j$ is 
\be
\left ( \begin{array}{c} j \\2b \end{array}\right )\frac{(2b)!}{b! 2^b}.
\ee
For the moments we thus find
(recall the notation (\ref{notation-n})):    
\be
\left\langle\left(n(\mu)-n(-\mu)\right)^j\right\rangle_{1+1}&=&
\sum_{b=0}^{{\rm Int}(j/2)} \left ( \begin{array}{c} j \\2b \end{array}\right )(2b-1)!!
[2(\chi_{ud}^B + \chi_{ud}^I)]^b(2\nu_I)^{j-2b}\nn \\
&=& \sum_{b=0 }^{{\rm Int}(j/2)}\left ( \begin{array}{c} j \\2b \end{array}\right ) 
(2\nu_I)^{j-2b}
\frac 1{\sqrt{\pi(\chi_{ud}^B + \chi_{ud}^I)}}
\int_{-\infty}^\infty du \ (2u)^{2b} e^{-u^2/(\chi_{ud}^B + \chi_{ud}^I)}\nn\\
&=&\frac 1{\sqrt{\pi(\chi_{ud}^B + \chi_{ud}^I)}}
\int_{-\infty}^\infty dx \ (2x)^{j} e^{-(x-\nu_I)^2/(\chi_{ud}^B +
  \chi_{ud}^I)} ,
\ee
where ${\rm  Int}(j/2)$ is the integer part of $j/2$.
These are the moments of a Gaussian distribution centered at $\nu_I$.
The distribution of the real part of the quark number is thus given by
\be
\label{PRenB}
P^{1+1}_{\Ree[n]}(x) 
&=& \frac{1}{\sqrt{\pi(\chi_{ud}^B+\chi_{ud}^I)}} e^{-{(x-\nu_I)^2}/{(\chi_{ud}^B+\chi_{ud}^I)}}.
\ee
We see that $P^{1+1}_{\Ree[n]}(x)$ is properly normalized. The expectation value of the
real part of the quark number equals the isospin number in the phase
quenched theory evaluated at $\mu$
\be\label{vevRenB1p1}
\langle \Ree[n]\rangle_{1+1} = \int_{-\infty}^\infty d x \  x P^{1+1}_{\Ree[n]}(x) = \nu_I,
\ee
and  the average of the square of the real part of the quark number is 
given by
\be
\langle (\Ree[n])^2 \rangle_{1+1} = \int_{-\infty}^\infty d x \  x^2 P^{1+1}_{\Ree[n]}(x) =
\nu_I^2+\frac{1}{2}(\chi_{ud}^B+\chi_{ud}^I). 
\ee

Note that fluctuations of the quark number in the real direction
vanish at $\mu=0$. The reason is that the width
of the distribution of the real part of $n$ goes to 0 for $\mu\to0$.

\subsection{The distribution of the imaginary part of the quark number}  
\label{subsec:PImnB1p1}

In this subsection we derive the distribution of the imaginary part of the quark number 
defined by  
\be
P^{1+1}_{\Imm[n]}(y)\equiv\left\langle
\delta\left(y+i\frac{1}{2}(n(\mu)+n(-\mu))\right)\right\rangle_{1+1},
\ee
within 1-loop chiral perturbation theory for $\mu<m_\pi/2$. 

\vspace{5mm}

As in the previous subsection the distribution is determined 
by its moments which can be shown by representing the $\delta$-function as an
integral 
\be
\label{kint2}
P^{1+1}_{\Imm[n]}(y) = \frac{1}{2\pi}\int_{-\infty}^\infty dk \
e^{-iyk}\left\langle
e^{\frac{k}{2}(n(\mu)+n(-\mu))}\right\rangle_{1+1}
\ee
and expanding the exponential  
\be
\label{expandTr2}
\left\langle e^{\frac{k}{2} (n(\mu)+n(-\mu)) }\right\rangle_{1+1}
&=& \sum_{j=0}^\infty \frac{1}{j!}
\left\langle\left(\frac{k}{2}(n(\mu)+n(-\mu))\right)^j\right\rangle_{1+1}.  
\ee
The $j$th power of $\Imm[n]$ can be computed by introducing $j$
fermionic replica quarks and $j$ bosonic replica quarks
\be
\left\langle\left(n(\mu)+n(-\mu)\right)^j\right\rangle_{1+1} 
&=& \left .
\lim_{n_i\to0}\frac{1}{n_1\cdots n_j}d_{\mu_1}\cdots d_{\mu_j} 
\left\langle\prod_{i=1}^j \frac{\det(D+\mu_i\gamma_0+m)^{n_i}}{\det(D-\mu_i\gamma_0+m)^{n_i}}\right\rangle_{1+1}
\right |_{\mu_i=\mu}.
\ee 
Note that the ratio of the two determinants makes up the phase factor,
$\det(D(\mu))/\det(D(-\mu))=\exp(2i\theta(\mu))$. From  
Eq.~(\ref{1}) we see that the phase indeed generates the imaginary part of
$n$ after differentiation with respect to  the chemical potential.   

The replicated generating function within one-loop chiral perturbation theory 
is given by
\be
&&\hspace{-2cm}\left\langle\prod_{i=1}^j
  \frac{\det(D+\mu_i\gamma_0+m)^{n_i}}{\det(D-\mu_i\gamma_0+m)^{n_i}} \right\rangle_{1+1} \nn\\
&=& \exp\left(\sum_{l\leq m=1}^j 2 n_ln_m (G_0(\mu_l,\mu_m)-G_0(-\mu_l,\mu_m))+\sum_{l=1}^j 2n_l(G_0(\mu_l,\mu)-G_0(-\mu_l,\mu)) \right),
\label{genFimnBdist}
\ee
where we used that $\mu<m_\pi/2$ so that 
pion condensates are absent in the generating function.

We now take $d_{\mu_1}\cdots d_{\mu_j}$ of the generating function,
evaluate it at $\mu_i=\mu$ and take the replica limit $n_i\to0$.
The only difference with the previous section is the minus sign in front 
of  $G_0(-\mu_l,\mu_m)$ and  $G_0(-\mu_l,\mu)$. We thus find the moments
\be
\left\langle\left(n(\mu)+n(-\mu)\right)^j\right\rangle_{1+1}&=&
\sum_{b=0}^{{\rm Int}(j/2)} \left ( \begin{array}{c} j \\2b \end{array}\right )(2b-1)!!
[2(\chi_{ud}^B -\chi_{ud}^I)]^b(-2\nu_I)^{j-2b}\nn \\
&=& \sum_{b=0}^{{\rm Int}(j/2)} \left ( \begin{array}{c} j \\2b \end{array}\right ) 
(-2\nu_I)^{j-2b}
\frac {(-1)^b}{\sqrt{\pi(\chi_{ud}^I - \chi_{ud}^B)}}
\int_{-\infty}^\infty du \ (2u)^{2b} e^{-u^2/(\chi_{ud}^I - \chi_{ud}^B)}\nn\\
&=&\frac 1{\sqrt{\pi(\chi_{ud}^I - \chi_{ud}^B})}
\int_{-\infty}^\infty dy \  (2 i y)^{j} e^{(iy+\nu_I)^2/(\chi_{ud}^I - \chi_{ud}^B)}
\ee
These are the moments of a Gaussian distribution centered at $i\nu_I$.
Notice that $\chi_{ud}^I - \chi_{ud}^B > 0$, which follows from
the explicit expression for the one-loop result (see Eq.~(\ref{G0})).
Since
\be
{\rm Im } \ n(\mu) = \frac 1{2i}(n(\mu) + n(-\mu)),
\ee
the distribution of the imaginary  part of the quark number is  given by
 \be
\label{PimnB}
P^{1+1}_{\Imm[n]}(y) 
&=& \frac{1}{\sqrt{\pi(\chi_{ud}^I-\chi_{ud}^B)}} e^{{(iy+\nu_I)^2}/{(\chi_{ud}^I-\chi_{ud}^B)}}.
\ee
Note that the distribution takes complex values, as could be expected because
of the phase of the fermion determinant.
 Moreover, we have that $P^{1+1}_{\Imm[n]}(y)$ is properly
normalized, that the expectation value of the
imaginary part of the quark number equals $i$ times the isospin
number in the phase quenched theory
\be
\langle \Imm[n]\rangle_{1+1} = \int_{-\infty}^\infty d y \  y P^{1+1}_{\Imm[n]}(y) = i\nu_I,
\ee
and finally, that the average of the square of the imaginary part of the quark number is given by
\be
\langle (\Imm[n])^2 \rangle_{1+1} = \int_{-\infty}^\infty d y \  y^2 P^{1+1}_{\Imm[n]}(y) =
-\nu_I^2+\frac{1}{2}(\chi_{ud}^I-\chi_{ud}^B). 
\ee
In contrast to the real part, the width of the fluctuations of the
imaginary part of $n$, $(\chi_{ud}^I-\chi_{ud}^B)^{1/2}$, remains non zero for
$\mu\to 0$. 

\subsection{The distribution of the quark number}

We finally turn to the distribution of the full quark number 
defined by  
\be
P^{1+1}_{n}(x,y)\equiv\left\langle
\delta\left(x-\frac{1}{2}(n(\mu)-n(-\mu))\right)\delta\left(y+i\frac{1}{2}(n(\mu)+n(-\mu))\right)\right\rangle_{1+1},
\ee
within 1-loop chiral perturbation theory for $\mu<m_\pi/2$. In this section
we show that this distribution factorizes into the  distribution of the
real and imaginary part of the quark number.

\vspace{5mm}
Factorization occurs if the moments factorize. This can be easily seen
by writing the probability distribution as an integral over the 
characteristic function, i.e.
\be
\label{kint4}
P^{1+1}_{n}(x,y) = \frac{1}{(2\pi)^2}\int_{-\infty}^\infty dk_x dk_y \
e^{-ixk_x}e^{-iyk_y}\left\langle e^{i\frac{k_x}{2}(n(\mu)-n(-\mu))} e^{\frac{k_y}{2}(n(\mu)+n(-\mu))}\right\rangle_{1+1}.
\ee
If the moments of the real and imaginary parts of the quark number
factorize, the expectation values of the exponents in between the
brackets will factorize.

The relevant moments follow from a replicated generating function as follows
\be
&&\left\langle(n(\mu)+n(-\mu))^j(n(\mu)-n(-\mu))^k\right\rangle_{1+1} \\
&=& \left .
\lim_{n_i\to0}\frac{1}{n_1\cdots n_{j+k}}d_{\mu_1}\cdots d_{\mu_{j+k}} 
\left\langle\prod_{l=1}^j \det(D+\mu_l\gamma_0+m)^{n_l}\det(D-\mu_l\gamma_0+m)^{n_l}
\prod_{l=j+1}^{j+k}\frac{\det(D+\mu_l\gamma_0+m)^{n_l}}{\det(D-\mu_l\gamma_0+m)^{n_l}}\right\rangle_{1+1}
\right |_{\mu_i=\mu}.\nn 
\ee 
When we compute this replicated generating function in one-loop chiral
perturbation theory a big simplification takes place: The contributions from
Goldstone particles with one quark from the first two determinants and the other from one
of the two determinants in the ratio exactly cancel.  The reason is that the 
one-loop contribution of such mixed fermionic Goldstone particles 
occurs with the same combinatorial factor as the mixed bosonic Goldstone
particles but with the opposite sign. This was first observed in \cite{LSV}
where it was formulated as the absence of 
correlations between the phase factor and the magnitude of the fermion
determinant to one loop order in chiral perturbation theory. 
Hence the moments  of the real and imaginary part of the quark number
factorize.

The probability distribution therefore factorizes as
\be
\label{PnB}
P^{1+1}_{n}(x,y) &=& 
P^{1+1}_{\Ree[n]}(x)P^{1+1}_{\Imm[n]}(y)\nn\\
 &=&\frac{1}{\pi\sqrt{(\chi_{ud}^I)^2-(\chi_{ud}^B)^2}}
e^{-(x-\nu_I)^2/(\chi_{ud}^I+\chi_{ud}^B)}e^{(iy+\nu_I)^2/(\chi_{ud}^I-\chi_{ud}^B)}.
\ee
To check this first main result let us first note that since the
 distributions of 
the real and imaginary parts of the quark number are normalized, also
the product is normalized.
Moreover, the expectation value of the quark number is zero
\be\label{vevnB1p1}
\langle n\rangle_{1+1} = \int dx dy \ (x+iy) P^{1+1}_{n}(x,y) =  \int dx \ x
P^{1+1}_{\Ree[n]}(x) + i \int dy \ y P^{1+1}_{\Imm[n]}(y) = \nu_I + i
i\nu_I = 0. 
\ee
We see that the total quark number (which necessarily is zero in
chiral perturbation theory) is obtained only after a detailed
cancellation between the contribution from the real part and the
imaginary part. Such a detailed cancellation also occurs when we compute
the average of $n^2$  
\be
\langle n^2 \rangle_{1+1} & = &  \int dx dy \ (x+iy)^2
P^{1+1}_{n}(x,y) \nn\\ 
 & = & 
\int dx \ x^2 P^{1+1}_{\Ree[n]}(x) -\int dy \ y^2 P^{1+1}_{\Imm[n]}(y) 
+2i\int dx \ x P^{1+1}_{\Ree[n]}(x)\int dy \ y P^{1+1}_{\Imm[n]}(y) \nn\\
& = & \nu_I^2+\frac{1}{2}(\chi_{ud}^I+\chi_{ud}^B)
-(-\nu_I^2+\frac{1}{2}(\chi_{ud}^I-\chi_{ud}^B))
+2i\nu_Ii\nu_I = \chi_{ud}^B. 
\ee
Note that, even though the distribution of the baryon
number depends on the isospin density and off-diagonal
susceptibility, these quantities drop out when evaluating
the moments of the quark number operator.

Finally, we note that since $\chi_{ud}^I+\chi_{ud}^B\to0$ for
$\mu\to0$ the quark number distribution becomes localized on the
imaginary axis for $\mu=0$. This is in perfect agreement with the fact
that the quark number operator is anti-hermitian for $\mu=0$
cf. Eq.~(\ref{nBherm}).

\section{The PQ distribution of the quark number operator}

In this section  we give the result for the partially quenched
distribution of the quark 
number. To derive this,  all we need to notice is how the terms that mix the
replica quarks and physical quarks contribute to the generating function.

{\sl Real part:} For the real part we
consider Eq. (\ref{genFrenBdist}) where the expectation value is now
taken in the 
$1+1^*$ theory. Since $G_0(\mu_l,\mu)$ is invariant under a change of the
sign of both chemical potentials we get exactly the same mixing between the
replica sector and the physical sector as before. Hence the final
answer for the distribution of the real part of $n$ is again the same
\be\label{PRenBPQ}
P^{PQ}_{\Ree[n]}(x) = P^{1+1}_{\Ree[n]}(x).
\ee

{\sl Imaginary part:} The generating function for the imaginary part
(\ref{genFimnBdist}) changes when we consider the partially quenched
case. This time the two physical flavors make up an absolute square of the
fermion determinant while the replica flavors makes up the phase factor. As
we have seen before there are no correlations between these two factors
within one-loop
chiral perturbation theory. The generating function for the imaginary part
of the quark number in the partially quenched case is thus given by
Eq.~(\ref{genFimnBdist}) but without the single sum which mixes the replica and
the physical sector. Hence there are no linear terms after differentiation
and the final result is obtained from Eq.~(\ref{PimnB}) by setting $\nu_I=0$
\be
P^{PQ}_{\Imm[n]}(y) = \frac{1}{\sqrt{\pi(\chi_{ud}^I-\chi_{ud}^B)}}
e^{-{y^2}/{(\chi_{ud}^I-\chi_{ud}^B)}}. 
\ee
  
As it should (since we take vev's in the $1+1^*$ theory) both distributions
are real and positive. The full distribution is again the product of these
two since the factorization only involves replicated flavors. Thus we find
\be
\label{PnBPQ}
P^{PQ}_{n}(x,y) & = & P^{PQ}_{\Ree[n]}(x)P^{PQ}_{\Imm[n]}(y)
\nn\\
  & = & 
\frac{1}{\pi\sqrt{(\chi_{ud}^I)^2-(\chi_{ud}^B)^2}}
e^{-{(x-\nu_I)^2}/{(\chi_{ud}^I+\chi_{ud}^B)}}
e^{{-y^2}/{(\chi_{ud}^I-\chi_{ud}^B)}}.  
\ee

As a cross check we see that 
\be
\langle n^{PQ} \rangle_{1+1^*} = \nu_I
\ee
and 
\be
\langle (n^{PQ})^2 \rangle_{1+1^*} = \chi_{ud}^B+\nu_I^2,
\ee
are in agreement with table \ref{tab:1}.

In section \ref{sec:numproblem} we make use of these results when
discussing the problems faced by numerical lattice QCD at $\mu\neq0$.

\section{The distribution of the chiral condensate}
\label{sec:Pqbarq}

In this section we derive the distribution of $\sigma$ and study how the
chiral condensate $\langle\sigma\rangle$ builds up. As was the case 
for the
quark number, the operator $\sigma$ is not hermitian 
\be
\label{hermqbarq}
\sigma(\mu)^* = \left(
\Tr\frac{1}{D+\mu\gamma_0+m}\right)^* 
=\Tr\frac{1}{D-\mu\gamma_0+m}= \sigma(-\mu).
\ee 
Therefore we  derive the distribution in the complex
$\sigma$ plane. We start by computing the distribution of the
real and of the imaginary part separately. The distribution of the full
chiral condensate then follows as the product of the two. This is
precisely the same which happened for the baryon density. In fact,
the derivation for $\sigma$ is almost identical to the one for the 
quark number.

\vspace{5mm}

To see that the derivation is analogous to that for the baryon
number let us start with the distribution of $\Ree[\sigma]$ and
derive the generating function. We evaluate the distribution 
\be
P^{1+1}_{\Ree[\sigma]}(x)\equiv\left\langle
\delta\left(x-\frac{1}{2}(\sigma(\mu)+\sigma(-\mu))\right)\right\rangle_{1+1}, 
\ee
for $\mu<m_\pi/2$ to one-loop order in chiral perturbation theory. 

\vspace{5mm}

The $\delta$-function can be  represented as an integral
\be
\label{kint3}
P^{1+1}_{\Ree[\sigma]}(x) = \frac{1}{2\pi}\int_{-\infty}^\infty dk \
e^{-ixk}\left\langle
e^{+i\frac{k}{2}(\sigma(\mu)+\sigma(-\mu))}\right\rangle_{1+1} 
\ee
and the exponential is expanded  
\be
\label{expandTrqbarq}
\left\langle e^{i\frac{k}{2} (\sigma(\mu)+\sigma(-\mu))
}\right\rangle_{1+1} 
&=& \sum_{j=0}^\infty \frac{1}{j!}
\left\langle\left(
i\frac{k}{2}(\sigma(\mu)+\sigma(-\mu))\right)^j
\right\rangle_{1+1} ,
\ee
so that the distribution follows from the moments.

The moments can again be expressed in terms of a replicated
generating function.
This time the replica index labels the masses $m_1,\ldots,m_j$, 
\be
&&\hspace{-2cm}\left\langle\left(
\sigma(\mu)+\sigma(-\mu)
\right)^j\right\rangle_{1+1} \\
&=&\left . 
\lim_{n_i\to0}\frac{1}{n_1\cdots n_j}d_{m_1}\cdots d_{m_j} 
\left\langle\prod_{i=1}^j 
\det(D+\mu\gamma_0+m_i)^{n_i}(D-\mu\gamma_0+m_i)^{n_i}
\right\rangle_{1+1}
\right  |_{m_i=m}.\nn 
\ee 
We now need to keep track of the mass dependence of the 
replicated generating function  
\be
&&\hspace{-3cm}\left\langle\prod_{i=1}^j
\det(D+\mu\gamma_0+m_i)^{n_i}(D-\mu\gamma_0+m_i)^{n_i}
\right\rangle_{1+1} 
\\
&=& \exp\left(\sum_{l\leq m=1}^j 2 n_ln_m
(G_0(\mu,\mu,m_l,m_m)+G_0(-\mu,\mu,m_l,m_m))\right. \nn\\
&& \hspace{1cm} \left.+\sum_{l=1}^j
2n_l\left[m_l\Sigma+(G_0(\mu,\mu,m,m_l)+G_0(-\mu,\mu,m,m_l))\right]
\right). \nn
\label{genFreqbarqdist}
\ee
Compared to the generating function for $\Ree[n]$ in
Eq.~(\ref{genFrenBdist}) there are two differences: {\sl 1)} The replica index
now labels the quark masses instead of the 
chemical potentials. {\sl 2)} The term linear in the replica
number contains  the mean field value, $\Sigma$, of the
chiral condensate. However, the entire structure of the generating
function remains and the combinatorics associated with the
differentiation  is identical to that for the quark number. Only
now the physical quantities appearing in the expressions are
the chiral condensates and the chiral susceptibilities. We get
\be
\label{PReqbarq}
P^{1+1}_{\Ree[\sigma]}(x) 
&=& \frac{1}{\sqrt{\pi(\chi_S^I+\chi_S^B)}} 
e^{-{(x-\Sigma-\delta\Sigma^B-\delta\Sigma^I)^2}/{(\chi_S^I+\chi_S^B)}},
\ee
where we used the definitions in (\ref{notation-m}).

Likewise the distribution for the imaginary part of $\sigma$
\be
P^{1+1}_{\Imm[\sigma]}(y)\equiv\left\langle
\delta\left(y+i\frac{1}{2}(\sigma(\mu)-\sigma(-\mu))\right)\right\rangle_{1+1},  
\ee
follows form that for the imaginary part of the quark number simply
by re-identifying the physical quantities which appear in the final
expression
\be
P^{1+1}_{\Imm[\sigma]}(y) = 
\frac{1}{\sqrt{\pi(\chi_S^I-\chi_S^B)}} 
e^{{(iy-\delta\Sigma^B+\delta\Sigma^I)^2}/{(\chi_S^I-\chi_S^B)}}.
\ee
Notice the absence of the mean field value of the chiral
condensate. In the generating function 
the mean field contribution from the fermionic replicas cancels
against the one from the replicated bosonic quarks. In agreement with
the hermiticity property (\ref{hermqbarq}) we see that the width of 
$P^{1+1}_{\Imm[\sigma]}$ vanishes for $\mu\to0$ while the width of
the distribution of the real part remains nonzero. From the explicit expression
for $G_0$ it can be easily shown that $\chi_S^I - \chi_S^B > 0$.

Since the structure of the generating function is unchanged the
full distribution,  
\be
P^{1+1}_{\sigma}(x,y)\equiv\left\langle
\delta\left(x-\frac{1}{2}(\sigma(\mu)+\sigma(-\mu))\right)
\delta\left(y+i\frac{1}{2}(\sigma(\mu)-\sigma(-\mu))\right)\right\rangle_{1+1},  
\ee
again factorizes
\be
P^{1+1}_{\sigma}(x,y)=P^{1+1}_{\Ree[\sigma]}(x)P^{1+1}_{\Imm[\sigma]}(y). 
\ee
Thus we have 
\be
P^{1+1}_{\sigma}(x,y) = \frac{1}{\pi\sqrt{(\chi_S^I)^2-(\chi_S^B)^2}} 
e^{-{(x-\Sigma-\delta\Sigma^B-\delta\Sigma^I)^2}/{(\chi_S^I+\chi_S^B)}}
e^{{(iy-\delta\Sigma^B+\delta\Sigma^I)^2}/{(\chi_S^I-\chi_S^B)}}.
\ee

As a check of this second main result we compute the chiral condensate
\be
\langle\sigma\rangle_{1+1} = \int_{-\infty}^\infty dx dy
\ (x+iy)P^{1+1}_{\sigma}(x,y) = \Sigma + 2 \delta\Sigma^B,
\ee
and the square of the condensate 
\be
\langle \sigma^2 \rangle_{1+1} = \int_{-\infty}^\infty dx dy
\ (x+iy)^2 P^{1+1}_{\sigma}(x,y) = \chi_S^B+(\Sigma + 2
\delta\Sigma^B)^2. 
\ee
We see that the dependence on $\Sigma^I$ and $\chi_S^I$ has canceled 
and the results obtained in  table \ref{tab:1} are reproduced. 
The cancellations take place in exactly the same manner as for the
baryon density. We conclude that in both cases the $\mu$ dependence of
the fluctuations are induced through a coupling to the isospin charge
of the pions. These strong fluctuations in the real part and in the
imaginary part combine and leave the physical observable independent
of the chemical potential.

\vspace{5mm}

Finally, we give also the partially quenched distribution of the
chiral condensate
\be
P_{\sigma}^{PQ}(x,y) = \frac{1}{\pi\sqrt{(\chi_S^I)^2-(\chi_S^B)^2}} 
e^{-{(x-\Sigma-\delta\Sigma^B-\delta\Sigma^I)^2}/{(\chi_S^I+\chi_S^B)}}
e^{-{y^2}/{(\chi_S^I-\chi_S^B)}}.
\ee
The partially quenched expectation values from table \ref{tab:1} 
\be
\langle\sigma\rangle_{PQ} = \Sigma + \delta\Sigma^B+\delta\Sigma^I,
\ee
and
\be
\langle \sigma^2 \rangle_{PQ} =
\chi_S^B+(\Sigma+\delta\Sigma^B+\delta\Sigma^I)^2  
\ee
follow from  Gaussian integrations over the distribution function.

We will use both the partially quenched results and  the full results
to explain how to deal with some of the numerical
problems encountered in lattice QCD at nonzero chemical potential.

\section{Numerical Lattice QCD at non zero $\mu$}
\label{sec:numproblem}

Above we have derived the full and partially quenched distributions of
the quark number operator for $\mu<m_\pi/2$. Here we discuss how these
results can be of use when measuring the quark number operator by numerical 
lattice QCD.

\vspace{2mm}

The main problem encountered in numerical lattice QCD at nonzero chemical
potential is that operators may acquire their expectation values by
virtue of extremely delicate cancellations caused by the complex
valued fermion determinant in the path integral.
To illustrate the problem let us look at the distribution of the quark
number operator, Eq.~(\ref{PnB}). Notice  that the
distribution itself takes complex values, i.e.~it is not a
probability measure. This feature is also shared by the unquenched
eigenvalue density of the Dirac operator \cite{O,AOSV} and the distribution
of the phase of the fermion determinant \cite{LSV}. These complex
oscillations are all important in order to separate the physics of 
nonzero baryon chemical potential \cite{OSV} from that at isospin
chemical potential. Common to the three examples mentioned above
is that the amplitude of the complex oscillations is exponentially
large in the volume. Moreover, in all three cases, one must integrate over
on the order of $V$ periods of the oscillations in order to obtain a reliable
value of the baryon density or chiral condensate. Let us illustrate this
explicitly using the results derived above.

The unquenched distribution of the quark number,
$P^{1+1}_{n}(x,y)$, takes complex values, and, as we have seen in 
Eq.~(\ref{vevnB1p1}), both the real part and the imaginary part contribute to the
baryon density. The magnitudes are equal, $\nu_I$, but the signs are
opposite so that the total quark number vanishes. Since the complex 
oscillations are associated with the imaginary part, $y$, of the
quark  number, $n=x+iy$, let us ask: How large should $y_{max}$ be in
order that  
\be
\label{ymax}
\int_{-y_{max}}^{y_{max}} dy \ iy P_{\Imm[n]}^{1+1}(y) \sim -\nu_I.
\ee
The answer is: 
\be
y_{max}^2-\nu_I^2\gg\chi_{ud}^I-\chi_{ud}^B
\ee
Since $\nu_I\sim V$ and $\chi_{ud}^I-\chi_{ud}^B\sim V$ we find that
$y_{max}$ has to be only slightly larger than $\nu_I$. As the period
of the oscillations is of order unity we conclude that we have to
include on the order of $V$ oscillations in the integral. 
This can be a hard task to control numerically unless we know the
analytical form of the distributions. The results presented here give this
form to leading order. There will be corrections to this form in order
to induce a nonzero baryon number, but one should still expect  
large cancellations between  contributions from the real part and the
imaginary part.   

\begin{figure}[t!]
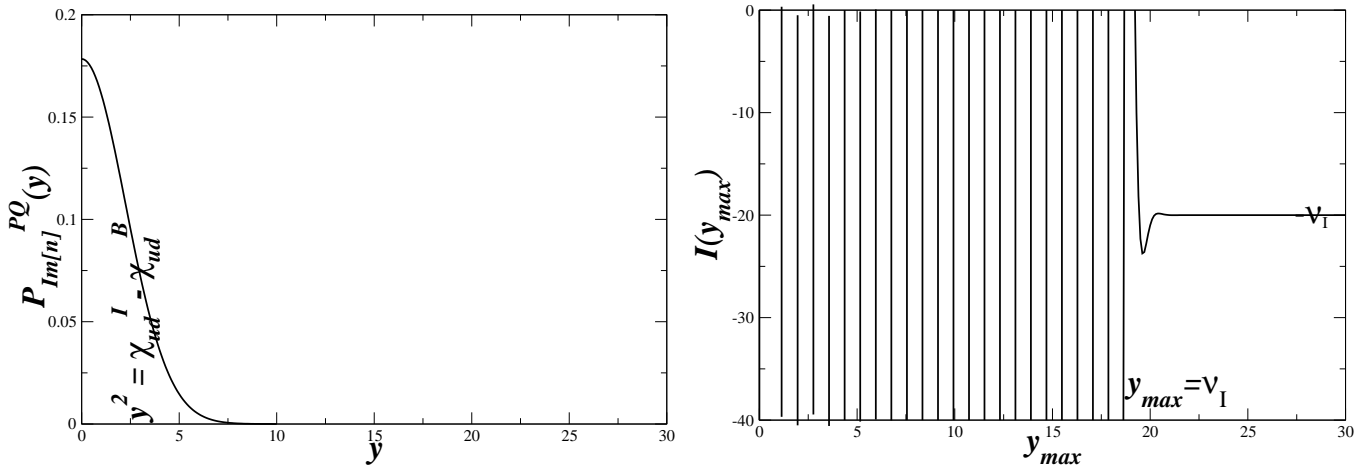

  \unitlength1.0cm
  \epsfig{file=PImPQ.eps,clip=,width=8.9cm}\hfill\epsfig{file=Iymax.eps,clip=,width=8.8cm}
  \caption{\label{fig:overlap} {\bf Left:} The distribution of
  the imaginary part of the quark number in the partially quenched
  simulation ensemble.  
  {\bf Right:} The contribution to the quark number in the 1+1
  target theory from the
  imaginary part of $n$ as a function of $y_{max}$, see
  Eq.~(\ref{ymax}). Only when $y_{max}$
  is larger than $\nu_I$ do we find that the contribution from the
  imaginary part cancels the
  one from the real part and leaves $\langle n\rangle_{1+1} = 0$. If
  $\nu_I\sim V$ is much larger than the width of the PQ
  distribution $\sqrt{\chi_{ud}^I-\chi_{ud}^B}\sim\sqrt{V}$, then there
  is a serious overlap problem. For the plot we have chosen $\nu_I=20$
  and $\chi_{ud}^I-\chi_{ud}^B=10$, and we already see that, 
 unless one also samples the far end of the tail, the
  configurations generated in the PQ simulation ensemble do  not have
  the values of $\Imm[n]$ required for the cancellation of the baryon
  number.}
\end{figure}

Not only can results from chiral perturbation theory help to understand the
detailed cancellations occurring in the integral, they can also 
give hints on which part of the integrand one needs to sample
numerically in lattice QCD:  
Imagine that we have generated an ensemble of configurations for the 
phase quenched weight. Then for the configurations in our simulation
ensemble, the quark number $n$ is distributed according to 
$P^{PQ}_{n}(x,y)$ as given in Eq.~(\ref{PnBPQ}).
Within the width of this distribution only on the order of $\sqrt{V}$
oscillations takes place. The reweighting from the simulation
ensemble to the full theory (target ensemble) therefore has to lift the
far tail of the phase quenched distribution. (For illustration, see
figure \ref{fig:overlap}).

Even though, the sign problem makes it hard to approach the
thermodynamic limit $V\to\infty$, it is worth to keep in mind that 
in a given numerical simulation we work with a finite volume where the
sign problem may be tractable depending on the value of $T$ and
$\mu$. We will address this issue in \cite{teflon}.

In the Complex Langevin \cite{Parisi,KW,FOC,AFP,aarts} approach to the QCD
sign problem the  
real and the imaginary part of the quark number are themselves
complex. In particular, this means that the imaginary part, $y$, is
not constrained to the real axis. Rather as we now show, $y$ will
fluctuate parallel to the real $y$ axis but shifted into the complex 
$y$ plane by $i\nu_I$. To show this we start from the complex action, 
\be
S = -\log[P_{\Imm[n]}^{1+1}(y)] = -(iy+\nu_I)^2/(\chi_{ud}^I-\chi_{ud}^B),
\ee
where we made use of the result in Eq.~(\ref{PimnB}). 
The flow equations for $y=a+ib$ are given by (the step size is
denoted by $\epsilon$)
\be
 a_{n+1} &=& a_n - \epsilon \Ree \left [\frac{dS}{dy} \right ]_{y=a_n+ib_n} +
 \sqrt{\epsilon} \eta_n \nn\\
         & = &  a_n - \epsilon \frac{2a_n}{\chi_{ud}^I-\chi_{ud}^B} +
 \sqrt{\epsilon} \eta_n 
\ee
and
\be
 b_{n+1} & = & b_n - \epsilon \Imm \left [\frac{dS}{dy} \right ]_{y=a_n+ib_n} \nn\\
         & =&  b_n - \epsilon \frac{2(b_n-\nu_I)}{\chi_{ud}^I-\chi_{ud}^B}.
\ee
Note that $a$ and $b$ decouple: $a$ fluctuates about zero while 
$b$ quickly moves to $\nu_I$ and stays there, since there is no noise
$\eta$ to kick it around. The complex Langevin algorithm, 
therefore, will essentially replace
\be
\int_{-\infty}^{\infty} y P_{\Imm[n]}^{1+1}(y)
\ee
by
\be
\int_{-a_{max}}^{a_{max}} da \ 
(a+i\nu_I) e^{-{a^2}/{(\chi_{ud}^I-\chi_{ud}^B)}},
\ee
where $a_{max}$ is the width of the region sampled by the
algorithm. Since the flow equation for $a$ is just that of the ordinary
real Gaussian, $\exp(-a^2/(\chi_{ud}^I-\chi_{ud}^B))$, and completely
decoupled from the imaginary part, the complex  Langevin algorithm
should not have any problems in sampling this. In other words,
$a_{max}$ is a good deal larger than $\sqrt{\chi_{ud}^I-\chi_{ud}^B}$
and the integral is close to $i\nu_I$ as desired. Again
there will be corrections to the one-loop results derived here and
complex Langevin must be able to take these into account correctly
in order to obtain the correct average quark number. However, as long as the
average quark number is much smaller than $\nu_I$ the fluctuations in
the complex $y$ plane should be expected to take place in the
neighborhood of $i\nu_I$. A similar example of a possible usefulness of
complex Langevin was presented in \cite{PhilippeLat2009}.

For purely imaginary values of the chemical potential the sign problem
is absent, and the quark number operator is  imaginary. 
In \ref{app:imagmu} we work 
out the distribution of the quark number for imaginary values of the
chemical potential. The result is a Gaussian
centered at zero. It is certainly true that simulations at imaginary $\mu$
\cite{owe1,maria,owe2} are easier -- the reason is the positivity of
the measure. This suggests that simulations at imaginary $\mu$ are as
easy as simulations at  $\mu=0$. Analytical and numerical 
studies show that the analytic continuation is well under control
at least for $\mu/T  < 1$ \cite{owe1,maria,owe2}, and it would be
interesting to interpret this result in the light of the distributions discussed
here. Analytic continuation to real $\mu$ has also been studied in
models without the sign problem \cite{continuation}.

\newpage

\section{Distributions for $\mu>m_\pi/2$}
\label{sec:mpi>2mu}

The chiral expansion of the generating functions relevant for the baryon
distribution take a different form when $\mu>m_\pi/2$. In this domain
the chemical potential induces a Bose condensate with a nonzero
isospin number even at the mean field level. The derivation presented
above is therefore not valid for $\mu>m_\pi/2$. In this section we
discuss the distributions of the quark number and the chiral
condensate for $\mu>m_\pi/2$.

\subsection{The fluctuations of the chiral condensate ($\mu>m_\pi/2$)}

For $\mu>m_\pi/2$ the quark mass is inside the spectrum 
of $D+\mu\gamma_0$
\cite{Gibbs,DominiqueJac,Factorization,AOSV}.  
Since an eigenvalue of $D+\mu\gamma_0$ 
can come very close to the quark mass the fluctuations of the
chiral condensate are much larger when $\mu>m_\pi/2$. In order to 
quantify the fluctuations of the chiral condensate let us consider the
moments of the real and the imaginary part of $\sigma$. To
start we consider the quenched case.  

{\sl The odd moments of $\Imm[\sigma]$:} Because the quenched
weight does not depend on the sign of $\mu$ the odd moments vanish  
\be
\langle (\sigma-\sigma^*)^{2p+1} \rangle = 0.
\ee
It follows that the quenched distribution of the imaginary part of the
chiral condensate must be an even function; that is:
$P^{(N_f=0)}_{\Imm[\sigma]}(y)$ is symmetric in $y$. 

\vspace{2mm}

{\sl The even moments of $\Imm[\sigma]$:} The even moments are
nonzero. In fact, as we now show, they are divergent. To see
this let us first consider the second moment of the imaginary part 
\be
\langle (\sigma-\sigma^*)^2 \rangle = 2\langle \sigma^2 \rangle 
- 2 \langle \sigma \sigma^* \rangle. 
\ee
If we express $\sigma$ in terms of the eigenvalues, $z_k$, of
$D+\mu\gamma_0$  
\be
\sigma = \sum_{k} \frac{1}{z_k+m}
\ee
we see that $\langle \sigma  \sigma^*
\rangle$ includes an absolute squared pole, which gives rise to
a logarithmic singularity. So we have
\be
\langle (\sigma-\sigma^*)^2 \rangle = -2 \langle
        \sigma  \sigma^* 
\rangle + {\cal O}(\epsilon^0).  
\ee  
Obviously, the logarithmic singularity is only present when the quark mass is
inside the support eigenvalue density. In fact, it can be shown
explicitly within chiral Random Matrix Theory that the
divergent part of the second moment is proportional to the eigenvalue
density evaluated at the quark mass
\be
\langle (\sigma-\sigma^*)^2 \rangle  
&=& -2 \left \langle \sum_k\frac 1{|z_k+m|^2} \right \rangle  
= -2 \int_{C_\epsilon} d^2 z \rho(z,z^*) 
\frac 1{|z+m|^2} \nn\\
&=& 4\pi \log(\epsilon) \rho_{N_f=0}(z=m, z^*=m)  
= \log(\epsilon)\theta(|\mu|-m_\pi/2)\frac{V\Sigma^2}{\mu^2F^2}.
\ee   
Here, $C_\epsilon$ is the complex plane excised by a sphere of radius 
$\epsilon$ centered at $-m$. The quenched eigenvalue density  
is the leading order result from chiral perturbation
theory  \cite{DominiqueJac}.   
We conclude that the quenched second moment of $\Imm[\sigma]$
is logarithmically divergent if the quark mass is inside the support
of the eigenvalue density, i.e.~if $\mu>m_\pi/2$.

The higher even moments of $\sigma$ have more severe
divergences. The most divergent term is the one with the same
powers of $\sigma$ and $\sigma^*$, so that,
\be
\langle (\sigma-\sigma^*)^{2p} \rangle 
\simeq (-1)^p\frac{(2p)!}{p! p!} \langle (\sigma \sigma^*)^p
\rangle.  
\ee 
In order to understand this better let us work out the details explicitly for
$p=2$. Again we keep only the leading divergence 
\be
\langle (\sigma-\sigma^*)^4 \rangle  
 & = & 6 \langle (\sigma \sigma^*)^2 \rangle \nn\\ 
 & = & 6\left\langle \sum_{g,h,j,k} 
\frac 1{z_g+m}\frac 1{z_h+m}\frac 1{z_j^*+m}\frac 1{z_k^*+m}
\right\rangle.
\ee
The most singular terms are the diagonal terms with $g=h=j=k$ 
which  lead to a $1/\epsilon^2$ singularity which
dominates the integral over the spectral density. 

Because the most divergent term is also given by the diagonal part
of the sum for $p>2$,
 it is always proportional to the eigenvalue density. We thus obtain 
the general relation (for $p=2,3,4,\ldots$)
\be
\langle (\sigma-\sigma^*)^{2p} \rangle 
\sim (-1)^p\frac{(2p)!}{p! p!} \frac{1}{2(p-1)\epsilon^{2(p-1)}}
\theta(|\mu|-m_\pi/2)\frac{V\Sigma^2}{4\pi\mu^2F^2}.
\label{singcond}
\ee

\vspace{2mm}

We conclude that
$P^{(N_f=0)}_{\Imm[\sigma]}(y)$ must be symmetric in $y$ in order that the odd
moments vanish, and it must have a $1/|y|^3$ tail 
to reproduce the observed singularities.
When the quark mass is outside the spectrum of the Dirac operator
the same argument applies, but now the spectral density at $z=m$
is exponentially suppressed, and the singular terms vanish in the
thermodynamic limit.
While the
singularities tell us about the tail of the distribution, the leading
divergent behavior of the moments does not contain sufficient
information to obtain  $P^{(N_f=0)}_{\Imm[\sigma]}(y)$  for
smaller $y$.   
We have verified the $1/|y|^3$ tail of $P^{(N_f=0)}_{\Imm[\sigma]}(y)$
by a numerical simulation of quenched chiral Random Matrix Theory.

\vspace{3mm}

So far we have considered the quenched case. As a first step toward
the fully unquenched theory let us consider the phase quenched theory.
In this case the absolute square of the fermion determinant in the
measure, $\prod_j|z_j^2+m^2|^2$, shifts the singularities. We now have
\be
\langle(\sigma-\sigma^*)^2\rangle_{1+1^*} \sim \epsilon^0,
\ee  
\be
\langle(\sigma-\sigma^*)^4\rangle_{1+1^*} \sim \log(\epsilon)
\ee
and
\be
\langle(\sigma-\sigma^*)^{2p}\rangle_{1+1^*} 
\sim \frac{1}{\epsilon^{2(p-2)}}
\ee  
for $p>2$. This implies that the far tail of the distribution
of $\Imm[\sigma]$
drops of like $1/|y|^5$. The distribution is again even since
the odd moments still vanish because the absolute square of the fermion
determinant does not depend  on the sign of the chemical potential. 

Finally, let us briefly look at the unquenched case. Let us for
simplicity take $N_f=1$. With a single determinant in the measure the
odd moments no longer vanishes. Moreover, the odd moments can now also
diverge. For example,  
\be
\langle (\sigma-\sigma^*)^3 \rangle_{N_f=1} =
-\langle \sigma^2  \sigma^* \rangle_{N_f=1} 
\sim \log(\epsilon).
\ee
Again the $\log(\epsilon)$ singularity is only  present when the
quark mass is inside the support of the spectral density, i.e. for
$\mu>m_\pi/2$.  
In general we have
\be
\langle (\sigma-\sigma^*)^{2p+1} \rangle_{N_f=1} \sim
\frac{1}{\epsilon^{2(p-1)}} 
\ee
for $p>1$. The diverging odd moments show that the unquenched
distribution of $\Imm[\sigma]$ is not an even function of $y$.  
We expect that the unquenched distribution of $\Imm[\sigma]$
takes complex values.  

Unfortunately, due to the divergences, we have not been able to derive
the full distribution of the imaginary part of the chiral condensate
for $\mu>m_\pi/2$. 

\vspace{2mm}

For the real part of $\sigma$ the main difference is that the
odd moments are nonzero. The leading divergent part of the even
moments is the same (up to a sign) 
as for the imaginary part. 


\subsection{Fluctuations of the quark number ($\mu>m_\pi/2$)}  

The general arguments which gave us information about the distribution
of the chiral condensate also apply to the quark number for
$\mu>m_\pi/2$. For $\mu>m_\pi/2$ we have that $\mu$ 
is inside the support of the spectrum of $\gamma_0(D+m)$
\cite{adam}, and this again leads to enhanced fluctuations. A rerun of
the general arguments presented above show that also
the quenched distribution of the real and of the imaginary part of the baryon
number has an inverse cubic tail. Rather than repeating the arguments
let us instead discuss the cancellations which insure that the total
baryon number remains zero in the unquenched case (as must be true in
chiral perturbation theory). 

Let us first consider the real and imaginary part of the quark number density
for the quenched or phase quenched case.
The imaginary part of the quark number is given by 
 \be
2 i \,\Imm[n] = {\rm Tr } \frac {\gamma_0}{ \gamma_0 (D+m) +\mu}
+ {\rm Tr } \frac {\gamma_0}{ \gamma_0 (D+m) -\mu} .
\ee
Because the spectrum of $\gamma_0(D+m) $ is reflection symmetric about
the imaginary axis, and the average spectrum for the quenched and 
phase quenched theory is reflection symmetric both about the real and 
the imaginary axis, we find that also for $\mu > m_\pi/2$
\be
 \langle \Imm[n] \rangle_{N_f =0}
= \langle \Imm[n] \rangle_{1+1^*}=0.
\ee
 The real part of the quark number is given by 
 \be
2 \Ree[n] = {\rm Tr } \frac {\gamma_0}{ \gamma_0 (D+m) +\mu}
- {\rm Tr } \frac {\gamma_0}{ \gamma_0 (D+m) -\mu} .
\ee
This is the isospin density with a nonvanishing expectation value 
in the quenched and phase quenched theory for $\mu>m_\pi/2$
because of pion condensation. At mean field level in Chiral
Perturbation Theory the isospin density is \cite{SS}
\be
\label{SSMF}
\nu_I = \langle \Ree[n]\rangle_{1+1^*} = 2 \mu F^2 [1-(\frac {m_\pi^2}{4\mu^2})^2].
\ee

We now consider the unquenched case with $N_f=2$. 
In the supersymmetric formalism, the generating function for 
$\Ree[n]$ is given by 
\be 
Z(\mu_1,\mu_2) = \left \langle \frac { {\det}(\gamma_0 (D+m)+\mu_2)
                { \det}(\gamma_0 (D+m)-\mu_2)
                 {\det}^{2}(\gamma_0 (D+m)+\mu_1) }
{{\det}(\gamma_0 (D+m)+\mu_1){\det}(\gamma_0 (D+m)-\mu_1) } \right \rangle 
\label{znf2} 
\ee   
with 
\be 
2\langle \Ree[n]\rangle_{1+1} =  \left . \frac d{d\mu_2} 
\log Z(\mu_1,\mu_2 )\right |_{\mu_1=\mu_2 = \mu}. 
\ee
The generating function undergoes a phase transition to a pion condensed phase
at $\mu = m_\pi/2$. It can be 
interpreted as the average phase factor at $\mu_1$ for the phase quenched
theory at $\mu_2$. Such averages where studied in \cite{SV-bos}. At the mean
field level the generating function factorizes as
\be
Z(\mu_1,\mu_2) =  
\frac {\langle  {\det}(\gamma_0 (D+m)+\mu_2)
                { \det}(\gamma_0 (D+m)-\mu_2)
                 {\det}^{2}(\gamma_0 (D+m)+\mu_1) \rangle }
{\langle {\det}(\gamma_0 (D+m)+\mu_1){\det}(\gamma_0 (D+m)-\mu_1)  \rangle} 
\label{znf3}
\ee

The imaginary part of the quark number for two flavors 
can also be obtained from the
generating function
\be
 2i \langle \Imm[n] \rangle_{1+1} =
\left . \frac d {d\mu_1} \right |_{\mu_1=\mu_2=\mu} \log Z(\mu_1,\mu_2)
\ee
with
 \be
Z(\mu_1, \mu_2)  =  \left \langle 
\frac {\det ( \gamma_0(D+m) +\mu_1) \det ( \gamma_0(D+m) +\mu_2) 
\det ( \gamma_0(D+m) -\mu_2)} {\det ( \gamma_0(D+m) -\mu_1) } \right \rangle.
\label{z31}
\ee
The partition function (\ref{z31}) is a phase-quenched average phase factor
which was studied in \cite{maria}. At the mean field level this partition function
factorizes as
\be
 Z(\mu_1, \mu_2)  
= 
\frac{ \langle {\det }^2( \gamma_0(D+m) +\mu_1) \det ( \gamma_0(D+m) +\mu_2) 
 \det ( \gamma_0(D+m) -\mu_2) \rangle } 
{\langle \det ( \gamma_0(D+m) -\mu_1)\det ( \gamma_0(D+m) +\mu_1)  \rangle}.
\label{z32} 
\ee

We observe that the real and imaginary part of the quark number are obtained
from the same generating function. 
The mean field analysis of partition function (\ref{znf3})
was outlined  in \cite{OSV-phase}. It is determined by the action
\be
S = \frac 14 F^2{\rm Tr} [U,B] [U^{-1}, B] - {\Tr} M \Sigma ( U + U^{-1})
\ee
with the baryon matrix and the mass matrix given by
\be
B = {\rm diag} (\mu_1, \mu_1, \mu_2, -\mu_2), \qquad M = {\rm
  diag}(m,m,m,m). 
\ee
The ansatz for the mean field can be written as \cite{OSV-phase}
\be
U = R^{-1}_{kl}(\beta_{kl}) R_{4k}(\alpha_k) R_{kl}(\beta_{kl})
\ee
with $k = 1, 2, 3$ and $l\ne k$. Here, $R_{pq}(\alpha)$ is a rotation in 
the $pq$ plane by angle $\alpha$. 
There is one important difference with \cite{OSV-phase} 
though. Because the quark masses are equal and the chemical potential
are put equal after differentiation, the integration over $\beta_k$ cannot
be done by a saddle point approximation but has to be performed exactly.

In total there are 6 different saddle
points. For $\mu_1= \mu_2$ the action of each of the saddle point is
the same and the dependence on $\beta_k$ cancels. Therefore, the integral over
$\beta_k$ has to be performed exactly for $\mu_1 \ne \mu_2$, whereas the $\alpha_k$ are determined
by the saddle point equation.

For the real part of the quark number we obtain:
\be
2\langle \Ree[n] \rangle_{1+1} = - (12+4)c \mu F^2 [\cos^2 \bar \alpha - 1],
\ee
where the first term (i.e. 12) originates from the rotation matrices that
mix $-\mu_2,\mu_2, \mu_1$ and the second term (i.e. 4) from the rotation
matrices that mix $-\mu_2,\mu_1, \mu_1$. The real constant $c$ is a
normalization factor. The solution of the saddle point
equation is given by \cite{SS}
\be
\cos \bar \alpha = \frac {m_\pi^2}{4\mu^2}.
\ee
With this we have 
\be
\langle \Ree[n] \rangle_{1+1} = 8 c \mu F^2 [1-(\frac {m_\pi^2}{4\mu^2})^2],
\ee
which is proportional to the isospin density in the Bose condensed phase of
the phase quenched theory at mean field level,cf. (\ref{SSMF}).

For the imaginary part of the quark number we obtain contributions
both from the numerator and the denominator of the generating function.
The contribution that originates from the numerator is given by
\be
(4+4) \mu F^2 [\cos^2 \bar \alpha - 1].
\ee
where again the first and second term correspond to 
rotation matrices that
mix $-\mu_2,\mu_2, \mu_1$ and 
that mix $-\mu_2,\mu_1, \mu_1$ in this order.
The contribution from the denominator is given by 
\be
-24c\mu F^2[\cos^2 \bar \alpha - 1].
\ee
Since the real and imaginary parts are obtained from the same generating
function, the constant $c$ in the expressions is the same.
For the imaginary part of the quark number we thus obtain 
\be
2i \langle \Imm[n] \rangle_{1+1} = 16 c \mu F^2[\cos^2 \bar \alpha - 1].
\ee
As must be true in chiral perturbation theory the sum of the real and
imaginary part of the quark number vanishes. Here we have shown this
by explicitly computing both contributions. As for $\mu<m_\pi/2$ we
have found that the two terms are proportional to the isospin
density in the phase quenched theory. For $\mu>m_\pi/2$, however, the
isospin density is far greater due to Bose condensation of pions.   

\section{Conclusions}

The distribution of the quark number operator and the chiral condensate for
Euclidean QCD at nonzero chemical potential has been derived to
leading order in chiral perturbation theory. 
As the two operators take on complex values,
the distributions are over the complex plane. 
Moreover,
because of the phase factor of the fermion determinant, the
distributions are not real and positive. 
We have shown
how the complex oscillations of the unquenched distributions lead to 
large cancellations when evaluating the baryon density and chiral
condensate. These cancellations give a direct insight into the problems
faced by numerical lattice QCD at nonzero chemical
potential. Of course the net contribution from pions to the average
baryon density is zero within chiral perturbation theory. Nevertheless,
pions contribute manifestly to the distribution of the quark
number and the chiral condensate and hence to the noise produced in
numerical lattice gauge simulations at nonzero chemical potential. 


Most of the results were derived for $\mu<m_\pi/2$ to one-loop
order in chiral perturbation theory. 
Then the distributions of the chiral condensate and the quark number
take a Gaussian form as one 
might expect from the central limit theorem. The unquenched
distribution of the imaginary part of these observables takes on complex values.
For example for the imaginary part of the quark number, 
this distribution is a Gaussian that
is shifted in the imaginary direction by an
amount proportional to the volume (the isospin number in the phase
quenched theory when evaluated at the same value of the chemical potential). 
This is of course not 
possible in an ordinary reweighting scheme and the final
results for the baryon density instead relies on detailed cancellations
after taking into account the far tail of the distribution. However, within
the Complex Langevin method the real and imaginary parts of the
baryon density are  complexified and the imaginary part could  fluctuate
around $i$ times a quantity of order the volume.  

For $\mu>m_\pi/2$ the distributions of the real and the
imaginary part of the quark number and chiral condensate develop a
power law tail. This extreme enhancement of the fluctuations is a direct
consequence of
the quark mass being inside the spectral support of the Dirac
operator, $D+\mu\gamma_0$, and the chemical potential being  inside the
support of $\gamma_0(D+m)$.       

The analytical results provided here may help in interpreting state of
the art lattice simulations at non zero chemical potential. They can
be used as benchmark for attempts to link the fluctuations of the baryon
number to the presence of the tricritical point. Since we have understood
 fluctuations produced the pions this may help in
optimizing and developing numerical approaches to reduce this source of 
noise. 

\vspace{2mm}

\noindent
{\bf Acknowledgments:}
We would like to thank P.H. Damgaard and B. Svetitsky for discussions. 
This work was supported  by U.S. DOE Grant No. DE-FG-88ER40388 (JV) and the 
Danish Natural Science Research Council (KS). 

\renewcommand{\thesection}{Appendix \Alph{section}}
\setcounter{section}{0}

\section{Imaginary chemical potential.}
\label{app:imagmu}

In this appendix we give the result for the distribution of the quark number
operator when evaluated at purely imaginary values of the chemical
potential. The fermion determinant
is real for imaginary chemical potential, and numerical simulations
are possible for an even number of 
flavors \cite{owe1,maria,owe2}. 
In this case the quark  number operator is anti-hermitian
even when $i\mu$ is nonzero
\be\label{inherm}
n(i\mu)^*=\left(\Tr\frac{\gamma_0}{D+i\mu\gamma_0+m}\right)^*
=-\Tr\frac{\gamma_0}{D+i\mu\gamma_0+m}=-n(i\mu).
\ee
The distribution is therefore one-dimensional
\be
P^{1+1,i\mu}_{n(i\mu)}(y)\equiv\left\langle
\delta\left(y+in(i\mu)\right)\right\rangle_{1+1,i\mu}.
\ee 
The derivation of the distribution   for $\mu<m_\pi/2$ to one-loop order
in chiral perturbation theory
is simpler than for real $\mu$ since now there are no terms 
$n(-i\mu)$ which can couple to the isospin charge of the pions. We find 
\be
P^{1+1,i\mu}_{n(i\mu)}(y) =
\frac{1}{\sqrt{\pi|\chi_{ud}^B|}}e^{-{y^2}/{|\chi_{ud}^B|}}. 
\ee
As must be true within chiral perturbation theory we find that 
\be
\left\langle n(i\mu)\right\rangle_{1+1,i\mu} = 0.
\ee
Note also that the width of the distribution is independent of
$\mu$. As $\chi_{ud}^B$ is extensive the distribution of the quark density becomes a
$\delta$-function at the origin in the thermodynamic limit. 
 
We also get 
\be
\left\langle n(i\mu)^2\right\rangle_{1+1,i\mu} = |\chi_{ud}^B|.
\ee

One can of course also consider the distributions of $n(i\mu)\pm
n(-i\mu)$ which will be the analytic continuations of the
distributions of the analytic part of the  real and imaginary parts of $n$.

\end{document}